%% file: draft_PRD.tex
\documentclass[twocolumn,showpacs,aps,prd]{revtex4-1}
\usepackage{epsfig}
\usepackage{graphicx}% Include figure files
\usepackage{dcolumn}% Align table columns on decimal point
\usepackage{bm}% bold math
\usepackage{ltablex,booktabs}
\usepackage{overpic}
\usepackage{subfigure}
\usepackage{float}
\usepackage{color}
\usepackage{amsmath}
\usepackage{mathcomp}
\usepackage{mathrsfs}
\usepackage{multirow}
\usepackage{rotating}
\usepackage{amssymb}
\usepackage{gensymb}
\usepackage{amsmath}
\usepackage{tabularx}
\usepackage{tikz-feynman}
\usepackage[bookmarksnumbered, pdfstartview=FitH,colorlinks,urlcolor=blue, citecolor=blue,linkcolor=blue] {hyperref}
\begin{document}
\normalsize
\parskip=5pt plus 1pt minus 1pt
%\linenumbers
\title{\boldmath Search for CP violation in $e^+e^- \to \psi(3770) \to D^0\bar{D}^0$ via $D \to K^0_S\pi^0$}
%\vspace{-1cm}
\newcommand{\BESIIIorcid}[1]{\href{https://orcid.org/#1}{\hspace*{0.1em}\raisebox{-0.45ex}{\includegraphics[width=1em]{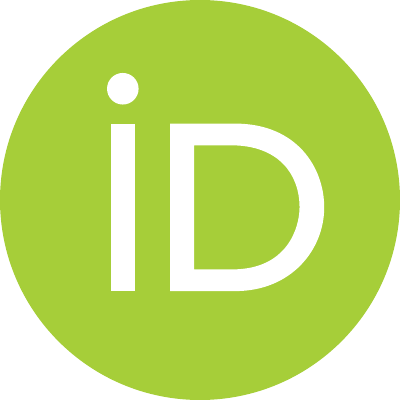}}}}

\input{author}

%\affiliation{}
%\vspace{-10cm}
%\date{\today}
%\setpagewiselinenumbers
\begin{abstract}
  Utilizing data sample of electron-positron
  collisions recorded with the BESIII detector at the center of mass energy
  of 3.773~GeV, corresponding to an integrated luminosity of
  20.28~fb$^{-1}$, we report the first search for the CP forbidden process $e^+e^- \to \psi(3773) \to D^0\bar{D}^0 \to (K^0_S\pi^0)(K^0_S\pi^0)$. No significant signal is observed. We set the upper limit on the observed cross section to be 7.37~fb, and the upper limit on the joint branching fraction of the C-odd correlated neutral $D$ pair $\mathcal{B}[(D^0\bar{D}^0)_{\text{C-odd}} \to (K^0_S\pi^0)(K^0_S\pi^0)] \textless 2.04 \times 10^{-6}$ at the 90\% confidence level.
\end{abstract}
%\pacs{}
\maketitle

%------------------------------------------------------------------------------
\section{Introduction}
The phenomenon of CP violation remains one of the least understood aspects of particle physics. Over the past sixty years, CP violation has been detected in several processes involving the $K$,  $B$ and $D$ meson systems. 
The recent observation of CP violation~\cite{prl-122-211803} in the charm quark sector is stimulating a wide discussion to understand its nature.
Identifying the sources of CP violation, whether within or beyond the Standard Model (SM), requires an accurate study performed across various processes.

\begin{figure}[htbp]
\begin{center}
	\includegraphics[width=3.5cm]{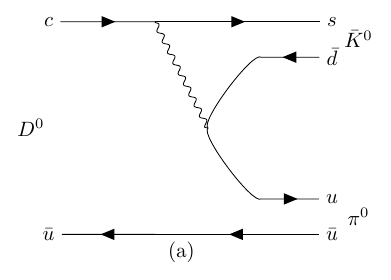}
	\includegraphics[width=3.5cm]{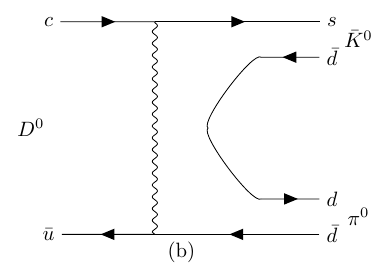}
	\includegraphics[width=3.5cm]{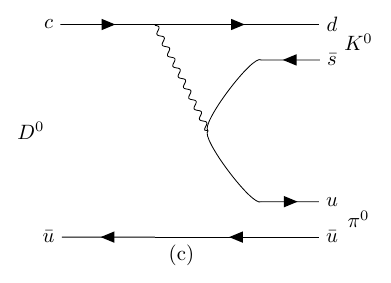}
	\includegraphics[width=3.5cm]{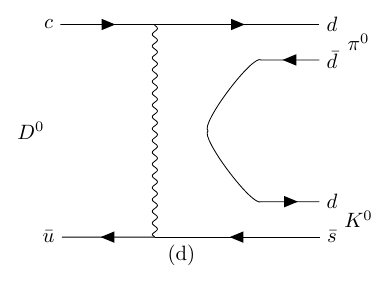}
\end{center}
\caption{Feynman diagrams for (a, b) Cabibbo-favoured process $D^0 \to \bar{K}^0 \pi^0$ and (c, d) doubly Cabibbo-suppressed $D^0 \to K^0 \pi^0$ at tree-level~\cite{prd-55-196218}.}
\label{fig:feynman}
\end{figure}

As suggested in Refs.~\cite{arxiv-2502.08907,plb-349-363,prd-55-196218},  a variety of neutral $D$ decays, including the semileptonic decays, the
hadronic CP and non-CP eigenstates, and the CP-forbidden states, provide an environment to obtain additional observations of CP violation, since they serve to clarify the ambiguities in current theoretical estimates and to shed some light on the dynamics of the $D^0-\bar{D}^0$ mixing~\cite{cpc-32-483}, of the subsequent  $K^0-\bar{K}^0$ mixing~\cite{jhep-12-011}, searching for possible sources of new physics beyond the SM. It is notable that the interference between these sources could potentially enhance the overall CP violation and affect the cross section~\cite{arxiv-2502.08907,prd-55-196218}, from which $D^0-\bar{D}^0$ mixing could be accessed. One possible hadronic CP eigenstate in $D^0$ decays is the $K^0_S\pi^0$ final state.  In the SM,  the decays $D^0\rightarrow \bar{K^0}\pi^0$ and
$D^0\rightarrow K^0\pi^0$ and their CP-conjugate processes are Cabibbo
favoured and doubly Cabibbo suppressed, respectively. Both
occur only through the tree-level quark diagrams, as illustrated in Fig.~\ref{fig:feynman}. The Kobayashi-Maskawa CP violating phase, interference between Cabibbo-favoured and doubly Cabibbo-suppressed diagrams, and the double-mixing from $D^0-\bar{D}^0$ and $K^0-\bar{K}^0$, generate rich opportunities for studying CP violation~\cite{arxiv-2502.08907,prd-110-l031301,lnp-591-237} and create an ideal platform for exploration.% In addition, $K^0$ regeneration effect may also contribute to the result~\cite{pr-124-1233}.

 Furthermore, the $D^{0}\bar{D}^{0}$ pairs produced in $e^+e^-$ annihilations originate from a virtual photon with quantum numbers $J^{PC}=1^{--}$, and while the two daughter $D$ mesons have spin-0, the initial virtual photon has spin-1. This results in the production of a P-wave two-meson system, which ensures that the two $D$ mesons must have opposite C $\times$ P. This is known as the Einstein-Podolsky-Rosen (EPR) correlation~\cite{CPViolation,pr-47-777780}.% The BESIII experiment has accumulated a large data sample corresponding to an integrated luminosity of $20.28$ fb$^{-1}$~\cite{cpc-48-123001} at the peak of the $\psi(3770)$ state, offering a unique opportunity to study the EPR correlation.

According to  CP conservation, Bose-Einstein statistics and the resulting EPR correlation, the process $e^+ e^- \to \psi(3770)\to D^0\bar{D^0}\rightarrow (K^0_S\pi^0) (K^0_S\pi^0)$ is forbidden since the final states share the same CP numbers, and thus violate the correlation. However, in the classical interpretation, the EPR correlation is entirely violated, with the $D$ mesons decaying independently with their own branching fractions (BFs). The expected number of produced events in the classical picture is given by
\begin{eqnarray}
  \begin{aligned}
    N_{\text{classical}} = N_{D^0\bar{D}^0} \mathcal{B}_{D^0} \mathcal{B}_{\bar{D}^0},
    \label{eq:crosssec_class}
  \end{aligned}
\end{eqnarray}
where $N_{D^0\bar{D}^0}$ is the total number of $D^0\bar{D}^0$ pairs produced in data, $\mathcal{B}_{D^0}$($\mathcal{B}_{\bar{D}^0}$)$=(1.240\pm0.022)\%$ is the BF of $D^0 \to K^0_S \pi^0$($\bar{D}^0 \to K^0_S \pi^0$)~\cite{prd-110-030001}. 

In this paper, we search for the $e^+ e^- \to \psi(3770)\rightarrow D^0\bar{D^0}\rightarrow (K^0_S\pi^0) (K^0_S\pi^0) $ process at $\sqrt{s}=3.773~$GeV.
We use the data sample collected by BESIII at $\sqrt{s}=3.773$ GeV, corresponding to an integrated luminosity of $20.28$ fb$^{-1}$~\cite{cpc-48-123001}. In the collected data sample, with Eq. \ref{eq:crosssec_class}, about 11000 events of $e^+ e^- \to \psi(3770)\to D^0\bar{D^0}\rightarrow (K^0_S\pi^0) (K^0_S\pi^0) $ process are expected from classical point of view~\cite{cpc-42-083001}, allowing to test the CP violation and the EPR correlation at $10^{-4}$ level.
We set an upper limit on the observed cross section and on the joint branching fraction of C-odd correlated neutral $D$ pair $\mathcal{B}[(D^0\bar{D}^0)_{\text{C-odd}} \to (K^0_S\pi^0)(K^0_S\pi^0)]$ at the 90\% confidence level (C. L.). Here, the subscript in $(D^0\bar{D}^0)_{\text{C-odd}}$ denotes the two neutral $D$ mesons are in C-odd correlation and have opposite C $\times$ P.
To avoid possible bias, a semi-blind analysis by using a sub-sample of the data, corresponding to an integrated luminosity of 2.93 fb$^{-1}$, is performed. Throughout this paper, charge conjugation is
always implied.

%------------------------------------------------------------------------------
\section{DETECTOR and DATA SETS} \label{sec:detector_dataset}
The BESIII detector~\cite{Ablikim:2009aa} records symmetric $e^+e^-$ collisions 
provided by the BEPCII storage ring~\cite{Yu:IPAC2016-TUYA01}
in the center-of-mass energy range from 1.84 to 4.95~GeV~,
with a peak luminosity of $1.1 \times 10^{33}\;\text{cm}^{-2}\text{s}^{-1}$ 
achieved at $\sqrt{s} = 3.773\;\text{GeV}$. 
BESIII has collected large data samples in this energy region~\cite{Ablikim:2019hff}. The cylindrical core of the BESIII detector covers 93\% of the full solid angle and consists of a helium-based
 multilayer drift chamber~(MDC), a plastic scintillator time-of-flight
system~(TOF), and a CsI(Tl) electromagnetic calorimeter~(EMC),
which are all enclosed in a superconducting solenoidal magnet
providing a 1.0~T magnetic field.
The solenoid is supported by an
octagonal flux-return yoke with resistive plate counter muon
identification modules interleaved with steel. 
%The acceptance of charged particles and photons is 93\% over $4\pi$ solid angle. 
The charged-particle momentum resolution at $1~{\rm GeV}/c$ is
$0.5\%$, and the 
${\rm d}E/{\rm d}x$
resolution is $6\%$ for electrons
from Bhabha scattering. The EMC measures photon energies with a
resolution of $2.5\%$ ($5\%$) at $1$~GeV in the barrel (end cap)
region. The time resolution in the TOF barrel region is 68~ps, while
that in the end cap region was 110~ps. The end cap TOF
system was upgraded in 2015 using multigap resistive plate chamber
technology, providing a time resolution of
60~ps,
which benefits 83\% of the data used in this analysis~\cite{etof}.

%------------------------------------------------------------------------------
Simulated Monte-Carlo~(MC) samples produced with {\sc geant4}-based~\cite{GEANT4}
software, which includes the geometric description of the BESIII detector and
the detector response, are used to determine the detection efficiency and to
estimate the background contributions. The simulation includes the beam energy
spread and initial state radiation~(ISR) in the $e^+e^-$ annihilation modeled
with the generator {\sc kkmc}~\cite{KKMC}. The inclusive MC samples, which are divided into 40 sub-samples with the same size as the data sample,
are used to study the background contributions. This sample includes the production of
open charm processes, the ISR production of vector charmonium(-like) states, and
the continuum processes incorporated in {\sc kkmc}. The known decay modes are
modeled with {\sc evtgen}~\cite{EVTGEN} using world averaged BF values~\cite{prd-110-030001},
and the remaining unknown decays from the charmonium states with
{\sc lundcharm}~\cite{LUNDCHARM}. Final state radiation from charged final
state particles is incorporated with {\sc photos}~\cite{PHOTOS}. The signal
detection efficiencies and signal shapes are obtained from signal MC simulation.

%------------------------------------------------------------------------------
\section{DATA ANALYSIS}
\label{chap:event_selection}
%------------------------------------------------------------------------------

In this analysis, $K^0_S$ and $\pi^0$ candidates are reconstructed via $K^0_S \to \pi^+\pi^-$ and $\pi^0 \to \gamma\gamma$ decays, respectively. The observed cross section $\sigma_{\text{obs}}$ of the process $e^+e^- \to \psi(3770) \to D^0\bar{D}^0 \to (K^0_S\pi^0)(K^0_S\pi^0)$ is given by
\begin{eqnarray}
  \begin{aligned}
    \sigma_{\text{obs}} = \frac{N_{\text{obs}}}{L \times \epsilon \times \mathcal{B}_{\text{sub}}},
    \label{eq:crosssec}
  \end{aligned}
\end{eqnarray}
and the joint branching fraction of the C-odd correlated $D^0\bar{D}^0$ is given by
  \begin{align}
    \mathcal{B}[(D^0\bar{D}^0)_{\text{C-odd}} \to (K^0_S\pi^0)(K^0_S\pi^0)] =&\notag \\ \frac{N_{\text{obs}}}{N_{D^0\bar{D}^0} \times \epsilon \times \mathcal{B}_{\text{sub}}}&, 
    \label{eq:jointbf}
  \end{align}
where $N_{\text{obs}}$ is the number of observed signal events,  $L$ is the total integrated luminosity of the data, $\epsilon$ is the efficiency for the event reconstruction, and $\mathcal{B}_{\text{sub}}$ is the BF of the sub-decay, given as $\mathcal{B}_{\text{sub}} = \mathcal{B}^{2}_{K^0_S} \times \mathcal{B}^{2}_{\pi^0}$, where $\mathcal{B}_{K^0_S}$ and $\mathcal{B}_{\pi^0}$ are the BFs of $K^0_S \to \pi^+\pi^-$ and $\pi^0 \to \gamma \gamma$, respectively.

The signal process
$e^+e^- \to \psi(3770) \to D^0\bar{D}^0 \to (K^0_S\pi^0)(K^0_S\pi^0)$ is searched using the double-tag method~\cite{prl-56-2140,arnps-73-285-314,nsr-8-11}. In this method, two $D$ mesons which decay through the $K^0_S\pi^0$ channel in the same event are simultaneously reconstructed. The final state $K^0_S\pi^0$ is reconstructed with the $K^0_S$ and $\pi^0$ candidates satisfying  the selections detailed below.

%------------------------------------------------------------------------------
Photon candidates are identified using isolated clusters  in the EMC. The EMC  time
is required to be within (0, 700)~ns from the event start time in order to
suppress fake photons due to electronic noise or $e^+e^-$ beam background. The deposited energy of each shower must be more than 25 MeV in the barrel region ($|\cos\theta|\textless 0.80$) and more than 50 MeV in the end cap region ($0.86 \textless |\cos\theta| \textless 0.92$), where
$\theta$ is the polar angle with respect to the $z$ axis, which is the symmetry
axis of the MDC.
The $\pi^0$ candidates are reconstructed through
the $\pi^0\to \gamma\gamma$ decay, with at least one
 photon in the barrel region. The two-photon invariant mass is required to be in the range
$(0.115, 0.150)$~GeV/$c^{2}$.
A 1C kinematic fit is performed constraining $M_{\gamma\gamma}$ to the
$\pi^{0}$ known mass~\cite{prd-110-030001}, and the $\chi^{2}$ of the fit must be less than 50. 
%The $\chi^{2}$ of the 1C kinematic fit constraining $M_{\gamma\gamma}$ to the
%$\pi^{0}$ known mass~\cite{prd-110-030001} must be less than 50. 

Charged tracks detected in the MDC are required to be within a polar angle range of $|\rm{cos\theta}|<0.93$. 
Each $K_{S}^0$ candidate is reconstructed from two oppositely charged tracks whose distance of closest approach to the interaction point along the $z$-axis is less than
20~cm. These tracks are assumed to be pions. They are constrained to originate from a common vertex and are required to have an invariant mass within (0.450, 0.550) GeV/$c^{2}$. The
decay length of the $K^0_S$ candidate is required to be greater than
twice the vertex resolution away from the IP. The quality of the vertex fit is ensured by requiring $\chi^{2} \textless 100$.

Signal $D$ meson pairs are reconstructed from two pairs of $K^0_S$ and $\pi^0$ mesons. For each $D$ meson candidate, we define the beam energy difference ($\Delta E$) as
\begin{eqnarray}
  \begin{aligned}
    \Delta E = E_{D} - E_{\text{beam}},
    \label{eq:de}
  \end{aligned}
\end{eqnarray}
where $E_D$ is the energy of the reconstructed $D$ meson and $E_{\text{beam}}$ is the beam energy, and we require $\Delta E$ to be within $(-0.042,~0.072)$ GeV. We define the distance from the left and right edges of the signal region to the most probable beam energy difference value, $\Delta E_{m}$, as 3$\sigma_L$ and 3$\sigma_R$, respectively, and define the $\Delta E$ significance~($\sigma_{\Delta E}$) for the $D$ meson as
\begin{eqnarray}
\begin{aligned}
\sigma_{\Delta E}=\left\{ \begin{array}{lr}
\frac{|\Delta E-\Delta E_{m}|}{\sigma_L}, & \Delta E \textless \Delta E_{m}, \\
\frac{|\Delta E-\Delta E_{m}|}{\sigma_R}, & \Delta E \textgreater \Delta E_{m},
\end{array}\right. .
\label{eq:desig}
\end{aligned}
\end{eqnarray}
For a candidate $D$ meson pair, the $\sigma_{\Delta E}$ of the two $D$ mesons are added up as ``total $\sigma_{\Delta E}$". For multiple candidate pairs, the one with the least ``total $\sigma_{\Delta E}$" is kept for further analysis.

\begin{figure}[htp]
  \begin{center}
    \includegraphics[width=0.40\textwidth]{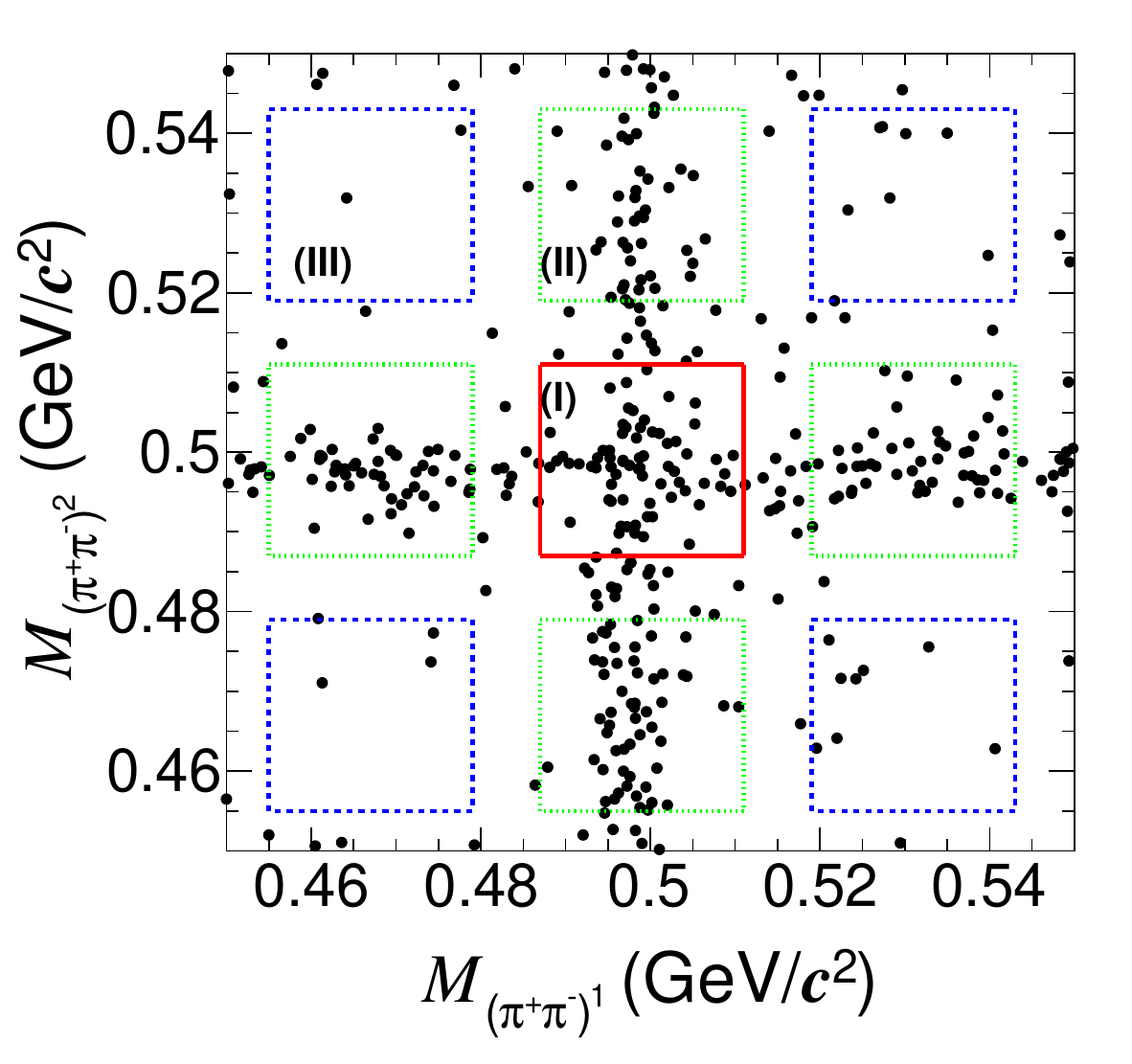}
    \caption{Two-dimensional $\pi^+\pi^-$ invariant mass distribution for the $K^0_S$ candidates from the two $D$ mesons in data (the order is arbitrary). The red solid box defines Region I, the green dotted box Region II, and the blue dashed box Region III.
    }
    \label{fig:MKs}
  \end{center}
\end{figure}

The dominant background is composed of events with $D^0(\bar{D}^0) \to \pi^+\pi^-\pi^0$ decays. To estimate this background, the invariant mass of the $K^0_S$ candidates are checked. Figure~\ref{fig:MKs} shows the two-dimensional $\pi^+\pi^-$ invariant mass distribution for both $D$ meson sides in the same event. We define the $K^0_S$ signal region as (0.487, 0.511) GeV$/c^2$, and the sideband regions as (0.455, 0.479) and (0.519, 0.543) GeV$/c^2$. We define Region I, corresponding to the signal region, where both $\pi^{+}\pi^{-}$ combinations lie in the $K^0_S$ signal region.  Region II is defined with either one of the $\pi^+\pi^-$ combinations in the $K^0_S$ signal region and the other in the $K^0_S$ sideband regions. The Region III has both the $\pi^+\pi^-$ combinations located in the $K^0_S$ sideband regions. Signal events are expected to peak in Region I, the $D^0 \to \pi^+\pi^-\pi^0$ background events are expected to distribute mainly in Region I and Region II, and other flat background events distribute in all the regions.

%and background in Region I can be estimated using events in Region II and Region III as $N_{\text{bkg}}=\frac{1}{2} N^{\text{(II)}} - \frac{1}{4} N^{\text{(III)}}$.

We use the beam-constrained mass $M_{\text{BC}}$ to identify $D$ mesons in the three regions, defined as
\begin{eqnarray}
\begin{aligned}
M_{\rm BC}=\sqrt{E^{2}_{\text{beam}}-|\vec{p}_{D}|^{2} c^{2}},
\end{aligned}
\end{eqnarray}
where $E_{\text{beam}}$ is the beam energy and $\vec{p}_{D}$ is the three-momentum of the reconstructed $D$ meson. 

\begin{figure*}[htbp]
  \begin{center}
    \includegraphics[width=0.32\textwidth]{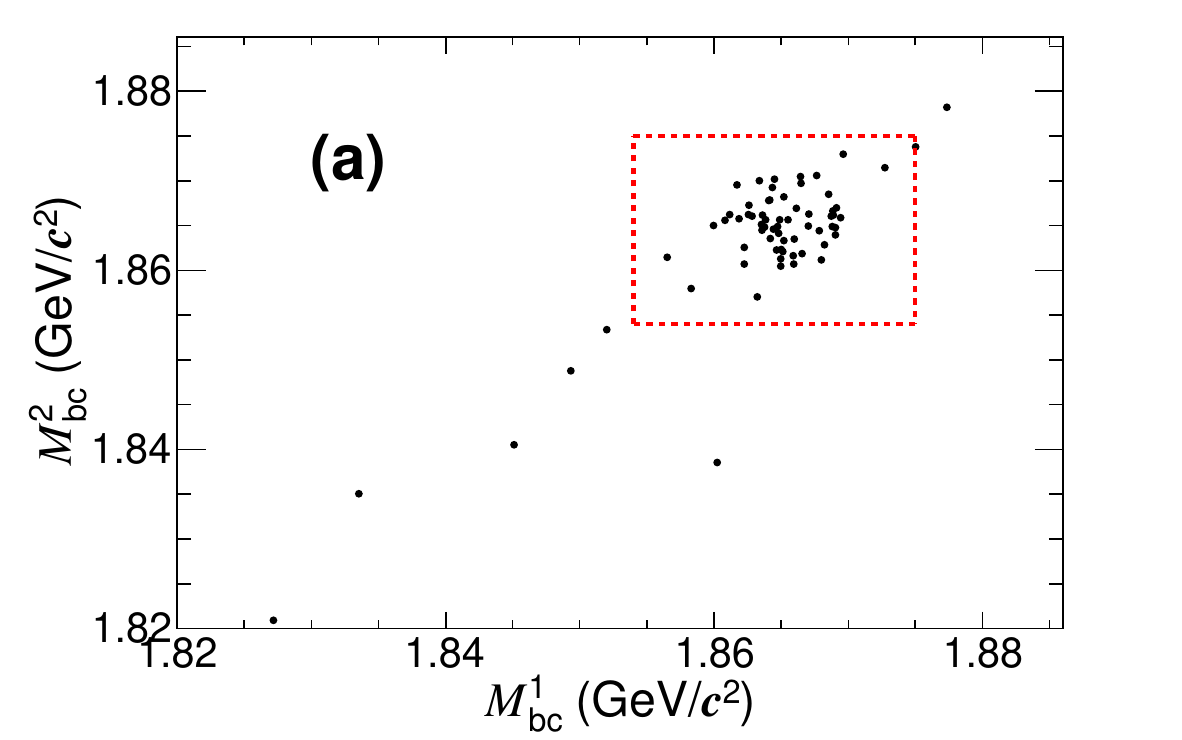}
    \includegraphics[width=0.32\textwidth]{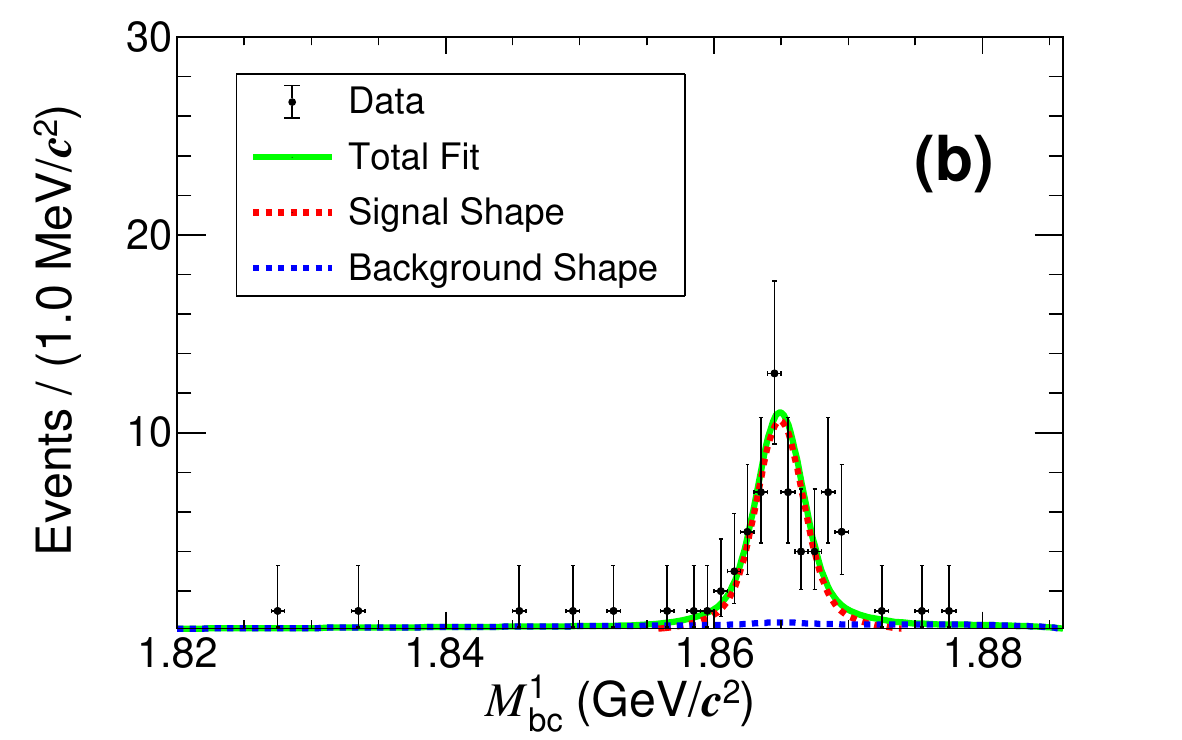}
    \includegraphics[width=0.32\textwidth]{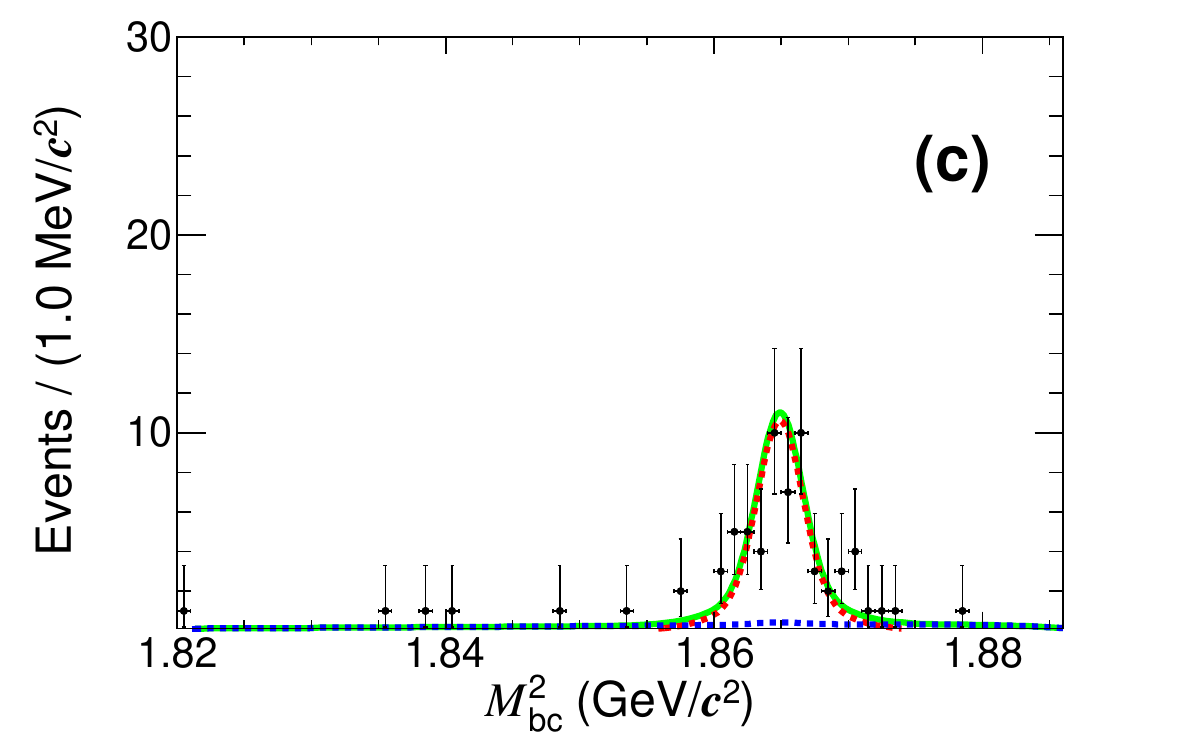}    
     \includegraphics[width=0.32\textwidth]{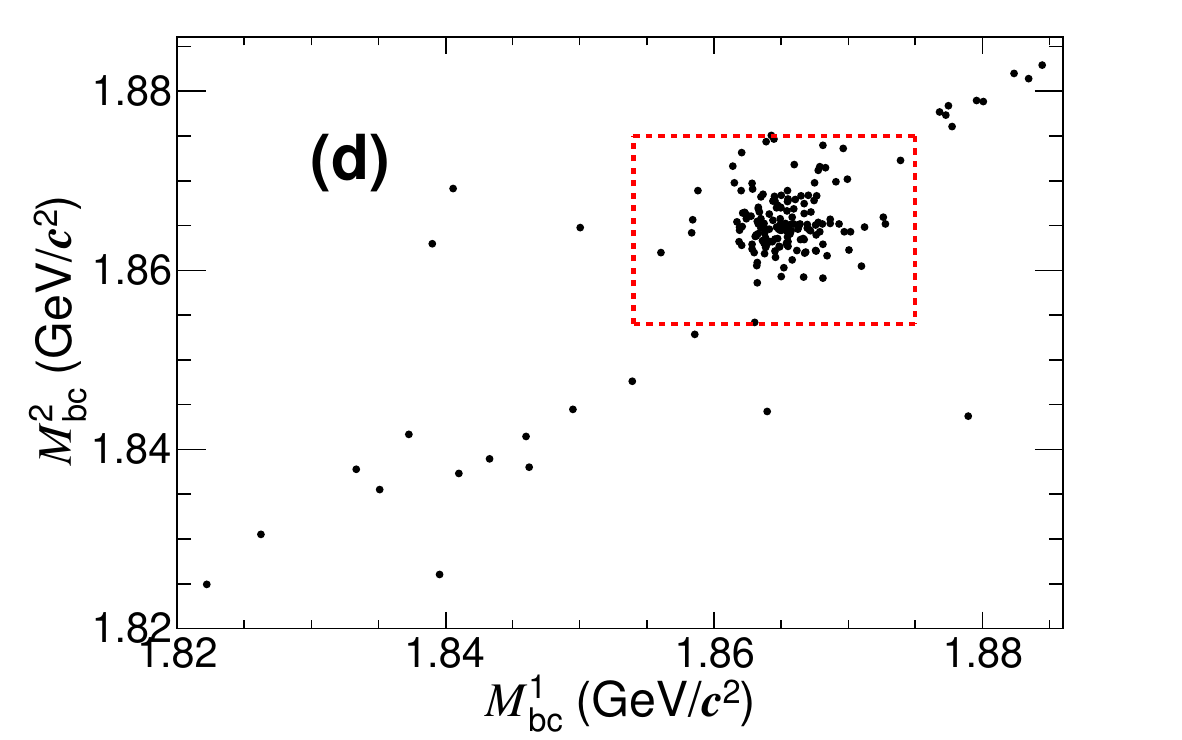}
    \includegraphics[width=0.32\textwidth]{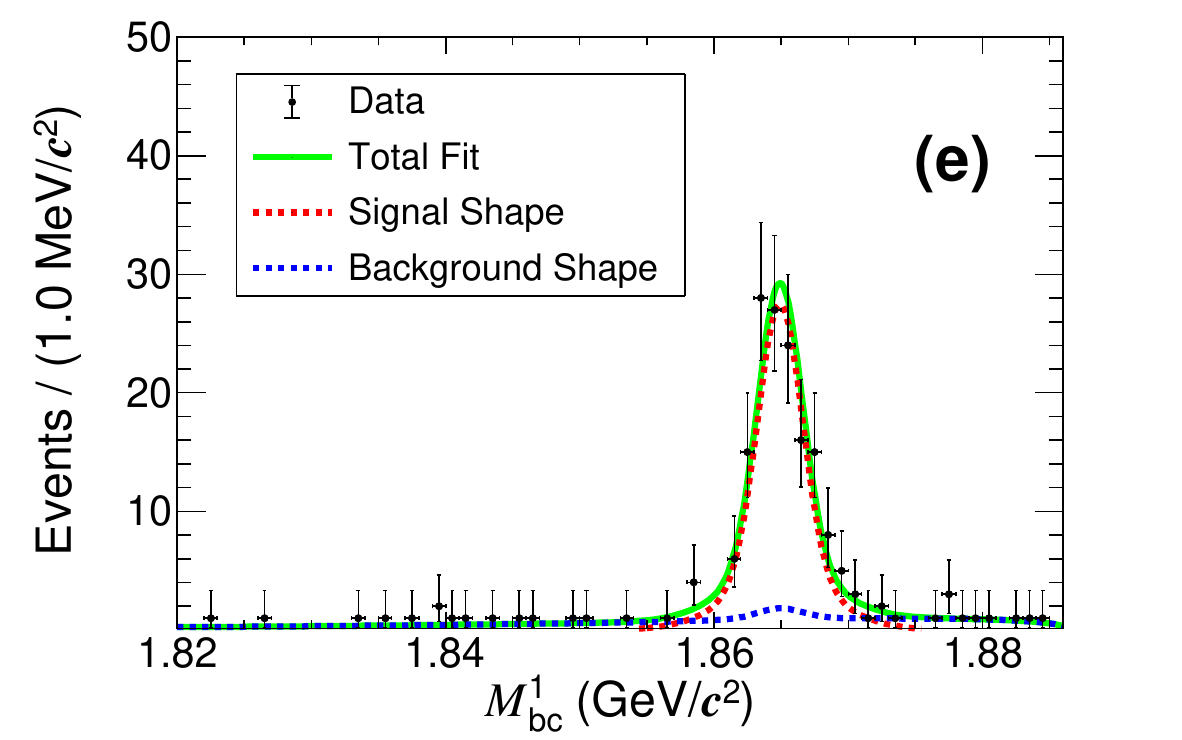}
    \includegraphics[width=0.32\textwidth]{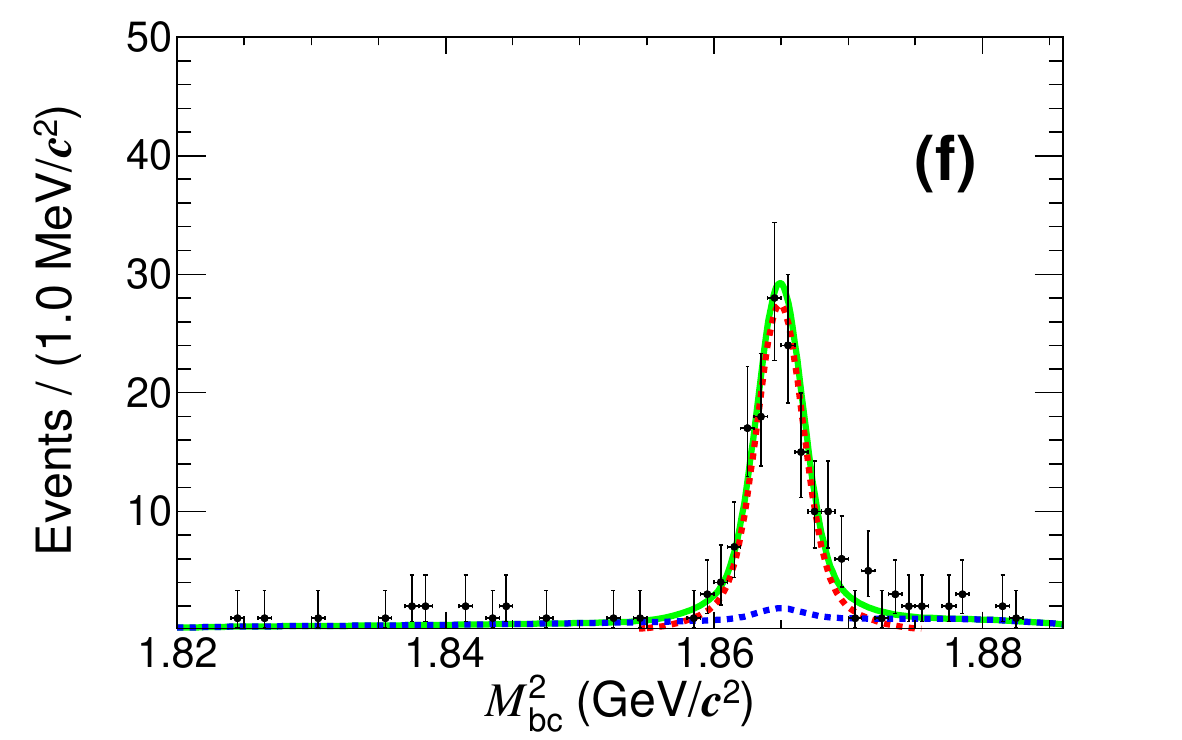}
     \includegraphics[width=0.32\textwidth]{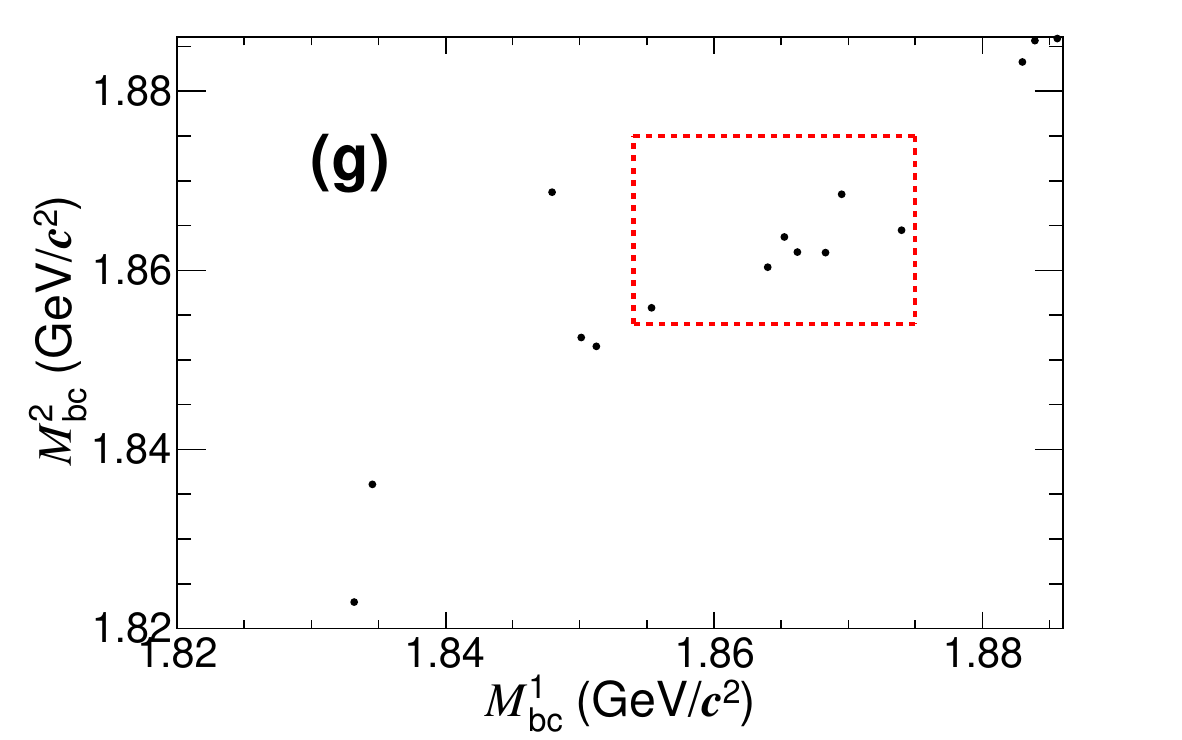}
    \includegraphics[width=0.32\textwidth]{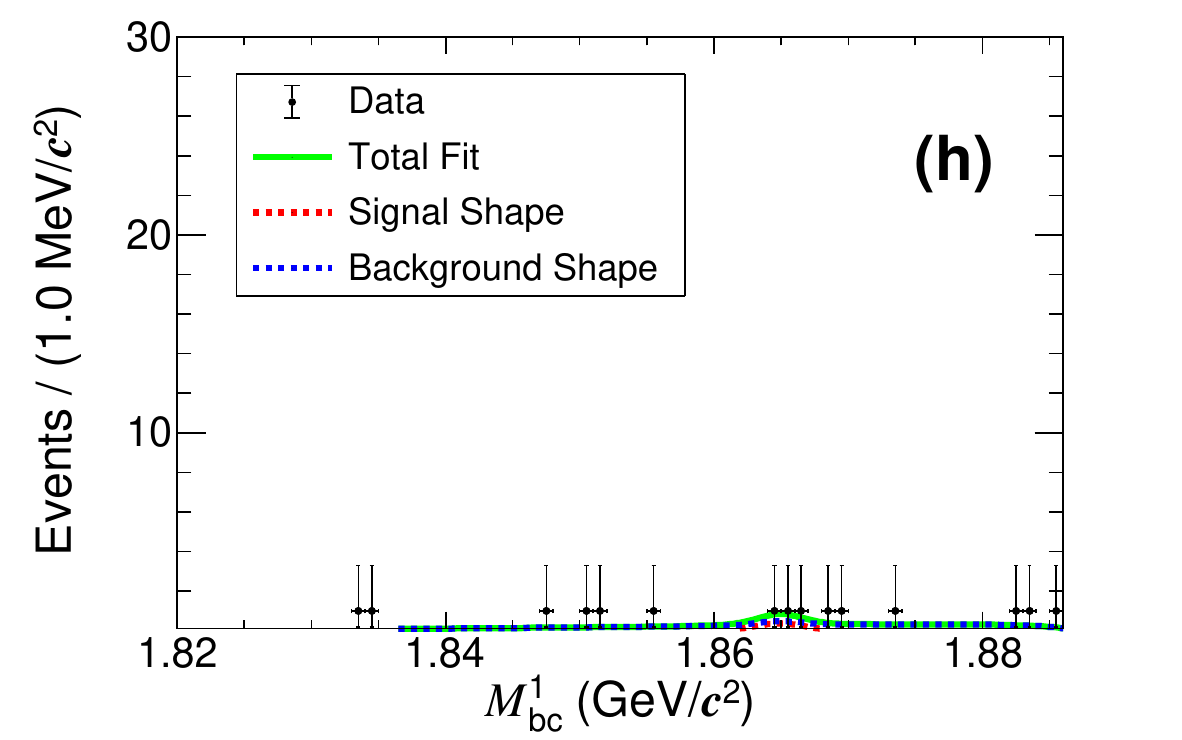}
    \includegraphics[width=0.32\textwidth]{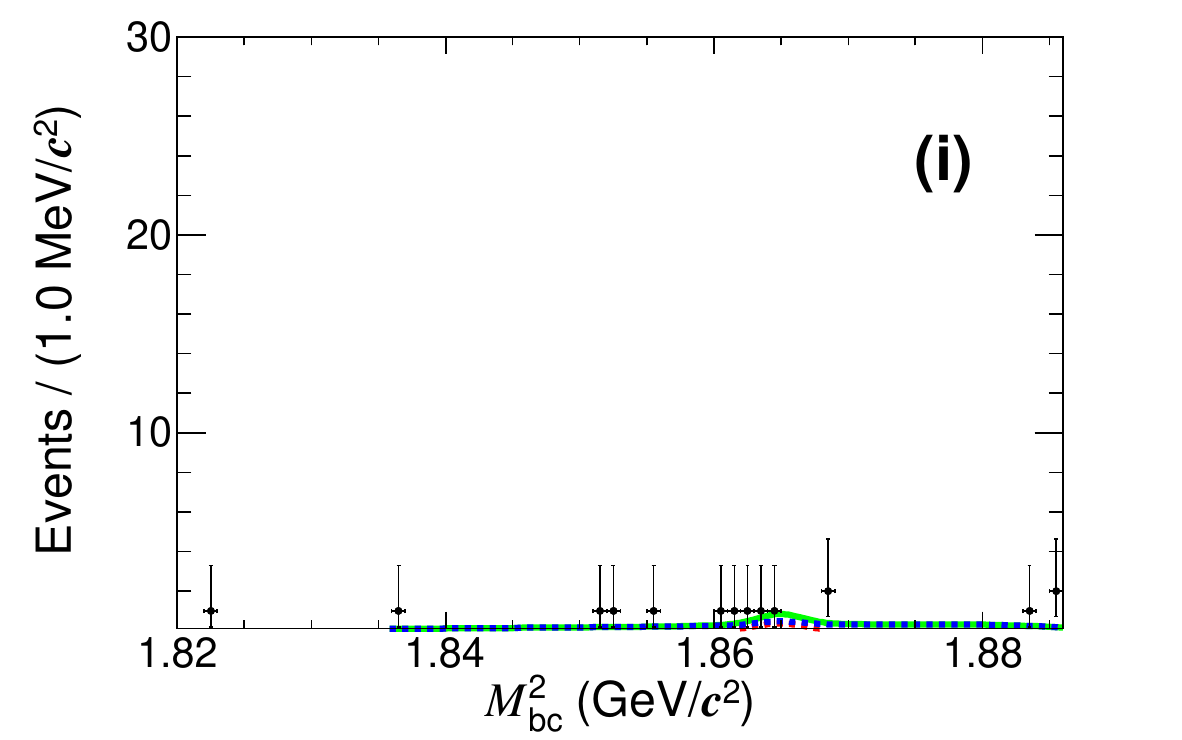}
    \caption{The $M_{\text{BC}}$ distributions of the reconstructed $D$ mesons. The first, second, and third rows correspond to events in Region I, II, and III, respectively. These Regions are shown in Figure~\ref{fig:MKs}; (a), (d), and (g) are the 2D distributions of $M_{\rm BC}$ for two $D$ mesons in the same event, where the red dashed squares refer to the $D$ signal region; (b), (e), and (h) are the $M_{\rm BC}$ distributions and fit result projections on one of the two $D$ meson, while (c), (f), and (i) are for the other $D$ meson. The order of $D$ mesons is arbitrary.
    }
    \label{fig:Mbc}
  \end{center}
\end{figure*}

The $D$ meson yield is acquired by a two-dimensional (2D) fit to the $M^{1}_{\text{BC}}$ versus $M^{2}_{\text{BC}}$ distribution of two $D$ mesons in the same event, where the order of the $D$ mesons is arbitrary. The probability density function (PDF) used in the fit comprises four components: ``Signal" describes candidates where both $D$ mesons are correctly reconstructed;  ``BKGI" and ``BKGII" denote the background where only one $D$ meson is correctly reconstructed;  ``BKGIII" describe the background that diagonally populates the 2D distributions.

Specifically, the four components are constructed as follows:\\
\indent Signal: $S_1 \times S_2$,\\
\indent BKGI: $S_1 \times f_2$,\\
\indent BKGII: $f_1 \times S_2$,\\
\indent BKGIII: $f_{1+2} \times S_{1-2}$.

Here, $1$ and $2$ correspond to the two dimensions of the 2D fit. The $S_i$ are Gaussian functions that describe the signal $D$ shape, while $f_i$ are ARGUS functions~\cite{argusfunc} that characterize the combinatorial background. The subscript $i$ denotes the argument of the functions.

%------------------------------------------------------------------------------

Finally, by fitting the $M_{\rm BC}$ distributions of the events within the three Regions defined in Fig.~\ref{fig:MKs}, three $D$ signal yields are obtained as $N^{\text{(I)}}=53\pm8$, $N^{\text{(II)}}=144\pm12$ and $N^{\text{(III)}}=2\pm2$. Notice that Region II (Region III) refers to all the green (blue) dashed square regions. The fit results are shown in Fig.~\ref{fig:Mbc}. Assuming the background is flat, the background in Region I can be estimated as $N_{\text{bkg}}=\frac{1}{2} N^{\text{(II)}} - \frac{1}{4} N^{\text{(III)}}=72\pm6$. The observed signal yield is determined to be $N_{\text{obs}}=N^{\text{(I)}} - N_{\text{bkg}}=-19 \pm 10$. The selection efficiency $\epsilon$ is estimated to be $(18.44\pm0.10)\%$ from signal MC sample. Since no significant signal is observed, an upper limit of $N_{\text{obs}}$ is determined with Bayesian approach, as shown in Fig.~\ref{fig:LL_smear_data}. The upper limit at the 90\% C. L. is 12.9, which is incorporated with the systematic uncertainties described in next section. This leads to an upper limit of observed cross section $\sigma_{\text{sig}}<7.37$ fb using Eq.~\ref{eq:crosssec}. The method to incorporate systematic uncertainty is discussed in the next section.

%------------------------------------------------------------------------------
\section{SYSTEMATIC UNCERTAINTY}
The systematic uncertainties on this measurement are summarized in 
Table~\ref{tab:Bf-syst-sum} and are classified into 
multiplicative terms $\sigma_{\epsilon}$, proportional to the measured BF, and additive terms $N_{\text{add}}$, independent of the measured BF.
\begin{table}[htp]
 \centering
%\begin{large}
\begin{tabular}{l|c}
\hline
\multicolumn{2}{c}{Multiplicative}\\
\hline
Source                                 & $\sigma_{\epsilon}$ (\%) \\
\hline
$K^0_S$ reconstruction                      & 1.5\\
$\pi^0$ reconstruction         & 1.7\\
$\Delta E$ signal region                           & 1.3\\
MC statistics                              & 0.2\\
Luminosity                                   & 0.5\\
Intermediate BF                          & 0.2\\
\hline
Total                                  & 2.7\\
\hline
\hline
\multicolumn{2}{c}{Additive}\\
\hline
Source & $N_{\text{add}}$ \\
\hline
Signal shape                           &3.5\\
Background shape                       &0.0\\
$N_{\text{obs}}$ estimation                                     & 10.3\\
\hline
Total                                  &10.9\\
\hline
\hline
\end{tabular}
%\end{large}
\caption{The estimated multiplicative and additive systematic uncertainties.}
\label{tab:Bf-syst-sum}
\end{table}
%------------------------------------------------------------------------------
The multiplicative uncertainties are from the efficiency determination and the
quoted BFs. The uncertainty from $K^0_S$ and $\pi^0$ reconstruction is studied from the control samples $D^{0} \to K^0_S \pi^{0}$, $D^{0} \to K^0_S \pi^{0} \pi^0$, $D^{0} \to K^0_S \pi^{0} \pi^+ \pi^-$ and $D^{0} \to K^{-} \pi^{+} \pi^{0}$, extracted from the same data sample~\cite{cpc-48-123001}. The uncertainty from the $\Delta E$ signal region selection is estimated by fitting the $\Delta E$ signal with a double-Gaussian function, and then using the $\pm3 \sigma$ range as alternative signal region;  the resulting change in $N_{\text{obs}}$ is taken as the related uncertainty. The uncertainty due to the size of the MC sample is assigned as  the statistical uncertainties related to the number of generated events.
The uncertainty for luminosity is quoted from Ref.~\cite{cpc-48-123001}, and the uncertainty for the intermediate BFs is quoted from PDG~\cite{prd-110-030001}. By adding these uncertainties in quadrature, the
total uncertainty $\sigma_{\epsilon}$ is estimated to be 2.7\%.
%------------------------------------------------------------------------------

The additive uncertainties directly affect the total number of observed signal events $N_{\text{obs}}$. The total additive uncertainty $N_{\text{add}}$ is estimated to be about 11 events. The uncertainty from the choice of the signal shape is obtained by altering the nominal signal
shape $S_i$ from a Gaussian function to a simulated signal shape, and the difference in $N_{\text{obs}}$ is considered as the related uncertainty. The uncertainty from the choice of the background shape is estimated in a similar way by changing the function $f_i$ with the simulated shape, and is found being negligible.  The uncertainty from the $N_{\text{obs}}$ estimation, including the background subtraction, is studied using forty inclusive MC samples with the same size of the data sample; the obtained values of $N_{\text{obs}}$ are fitted with a Gaussian function and we obtain as mean value  $0.3\pm1.2$ and as standard deviation  $7.7\pm0.9$. To be conservative, we consider the sum of the mean value and standard deviation, together with their uncertainties, as the uncertainty from the $N_{\text{obs}}$ estimation.
%------------------------------------------------------------------------------
\section{Results} 
\begin{figure}[htbp]
  \begin{center}
    \includegraphics[width=0.40\textwidth]{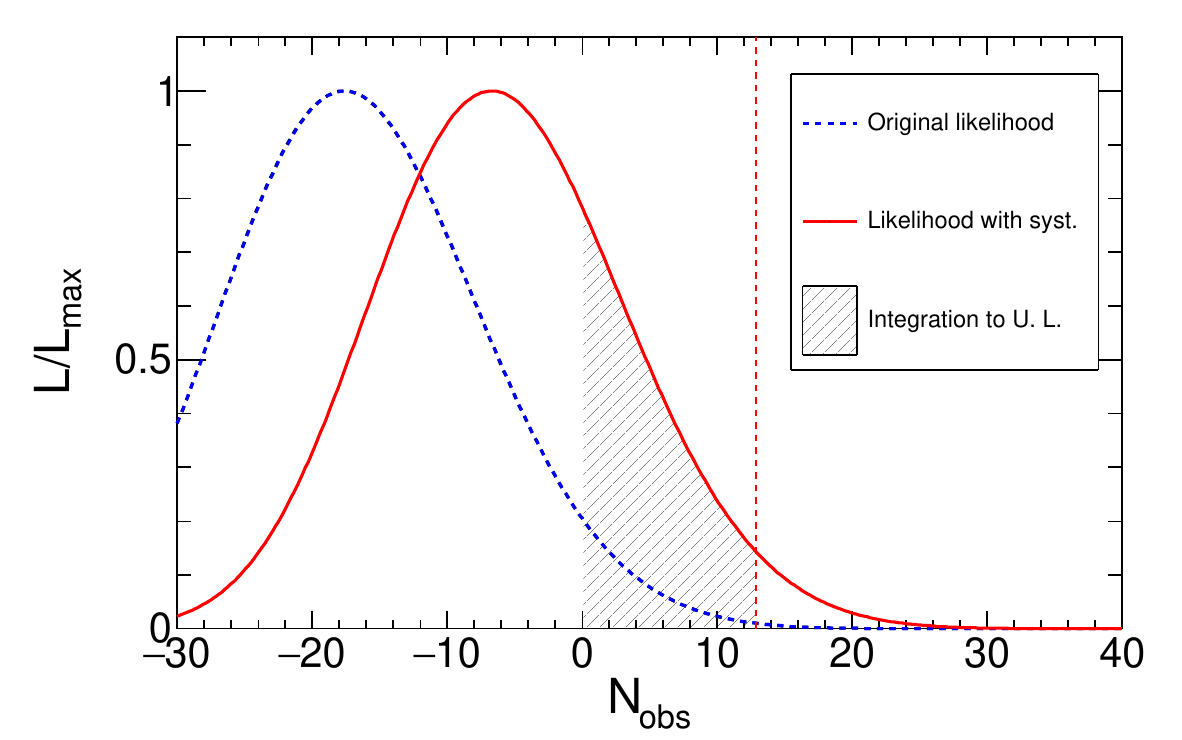}
    \caption{
      Likelihood distributions as a function of $N_{\text{obs}}$ of the data sample. The
      results obtained with and without incorporating the systematic
      uncertainties are shown as red solid and blue dashed curves,
      respectively. The dashed red line shows the result corresponding to the
      90\% C.L.. A flat prior for $N_{\text{obs}}$ from 0 to infinity is assumed. The shaded area shows the integration above zero to the upper limit.
    }
    \label{fig:LL_smear_data}
  \end{center}
\end{figure}

The multiplicative systematic uncertainty is incorporated in the calculation of
the upper limit following the prescriptions of Refs.~\cite{Stenson, CPC-39-113001}, as
\begin{eqnarray}
  \begin{aligned}
    L\left(N_{\text{obs}}\right)\propto \int^1_0 L\left(N_{\text{obs}}\frac{\epsilon'}{\epsilon}\right){\rm exp}\left[\frac{-\left(\epsilon'/\epsilon-1\right)^2}{2 (\sigma_{\epsilon})^2 }\right]d\epsilon' \,,
  \end{aligned}
\end{eqnarray}
where $L(N_{\text{obs}})$ is the likelihood distribution as a function of the assumed
observed signal number $N_{\text{obs}}$; $\epsilon'$ is the expected efficiency and $\epsilon$ is the MC-estimated efficiency. To take into account the additive systematic
uncertainty, the updated likelihood is then shifted upward according to the number of additive uncertainty events $N_{\text{add}}$. The blue dashed and red solid
curves in Fig.~\ref{fig:LL_smear_data} show the original and the modified likelihood
distributions, respectively.

The upper limit on $N_{\text{obs}}$ at the $90\%$
C.L., obtained by integrating the
resulting curve from zero to $90\%$, is $N_{\text{obs}}<12.9$. It gives an upper limit of the observed cross section $\sigma_{\text{obs}}(e^+ e^- \to \psi(3770)\rightarrow D^0\bar{D^0}\rightarrow (K^0_S\pi^0) (K^0_S\pi^0) )<7.37~$fb, and of the joint branching fraction $\mathcal{B}[(D^0\bar{D}^0)_{\text{C-odd}} \to (K^0_S\pi^0)(K^0_S\pi^0)]<2.04 \times 10^{-6}$ at $\sqrt{s}=3.773~$GeV.

%------------------------------------------------------------------------------
\section{Conclusion} \label{CONLUSION}
Using data taken at $\sqrt{s} = $ 3.773~GeV corresponding to an integrated luminosity of 20.28~fb$^{-1}$,
recorded by the BESIII detector, we perform the first search for
$e^+ e^- \to D^0\bar{D^0}\to (K^0_S\pi^0) (K^0_S\pi^0)$ following a semi-blind analysis. The quantum-correlated process
$e^+ e^- \to D^0\bar{D^0}\to (K^0_S\pi^0) (K^0_S\pi^0)$ is forbidden by CP conservation, and only allowed via $D^0-\bar{D}^0$ mixing and $K^0-\bar{K}^0$ mixing mechanism in the secondary decays, and interference between them. In this analysis, we establish the upper limit on the observed cross section
$\sigma_{\text{obs}}(e^+ e^- \to D^0\bar{D^0}\to (K^0_S\pi^0) (K^0_S\pi^0))<7.37$~fb; additionally, the joint branching fraction is constrained to $\mathcal{B}[(D^0\bar{D}^0)_{\text{C-odd}} \to (K^0_S\pi^0)(K^0_S\pi^0)]<2.04 \times 10^{-6}$, both at the 90\%
C.L. and at $\sqrt{s}=3.773~$GeV. This is the first exploration on the joint branching fraction and cross section of EPR correlated neutral $D$ pair. This method can be generalized to study other EPR correlated systems like $e^+e^- \to D^0\bar{D}^{*0}$, and the result provides crucial information for the CP violation of neutral $D$ meson and helps eliminate or set theoretical boundaries to potential new physics effect in charm physics~\cite{prd-55-196218}. In the future, by studying the interference between different CP violation sources and existing CP violation results in neutral $K$ mesons, more accurate cross section measurements will be able to bring a deeper insight into the $D^0-\bar{D}^0$ mixing parameters, and finally reveal the nature of this broken symmetry.

\begin{acknowledgements}
\label{sec:acknowledgement}
\vspace{-0.4cm}
The BESIII Collaboration thanks the staff of BEPCII (https://cstr.cn/31109.02.BEPC) and the IHEP computing center for their strong support. This work is supported in part by National Key R\&D Program of China under Contracts Nos. 2023YFA1606000, 2023YFA1606704; National Natural Science Foundation of China (NSFC) under Contracts Nos. 11635010, 11935015, 11935016, 11935018, 12025502, 12035009, 12035013, 12061131003, 12192260, 12192261, 12192262, 12192263, 12192264, 12192265, 12221005, 12225509, 12235017, 12361141819; the Chinese Academy of Sciences (CAS) Large-Scale Scientific Facility Program; CAS under Contract No. YSBR-101; 100 Talents Program of CAS; The Institute of Nuclear and Particle Physics (INPAC) and Shanghai Key Laboratory for Particle Physics and Cosmology; German Research Foundation DFG under Contract No. FOR5327; Istituto Nazionale di Fisica Nucleare, Italy; Knut and Alice Wallenberg Foundation under Contracts Nos. 2021.0174, 2021.0299; Ministry of Development of Turkey under Contract No. DPT2006K-120470; National Research Foundation of Korea under Contract No. NRF-2022R1A2C1092335; National Science and Technology fund of Mongolia; Polish National Science Centre under Contract No. 2024/53/B/ST2/00975; Swedish Research Council under Contract No. 2019.04595; U. S. Department of Energy under Contract No. DE-FG02-05ER41374
\end{acknowledgements}
%\section{Appendix A: }

\end{document}

%% file: author.tex
%\author{Author list}
\author{
\begin{small}
\begin{center}
M.~Ablikim$^{1}$\BESIIIorcid{0000-0002-3935-619X},
M.~N.~Achasov$^{4,b}$\BESIIIorcid{0000-0002-9400-8622},
P.~Adlarson$^{77}$\BESIIIorcid{0000-0001-6280-3851},
X.~C.~Ai$^{82}$\BESIIIorcid{0000-0003-3856-2415},
R.~Aliberti$^{36}$\BESIIIorcid{0000-0003-3500-4012},
A.~Amoroso$^{76A,76C}$\BESIIIorcid{0000-0002-3095-8610},
Q.~An$^{73,59,\dagger}$,
Y.~Bai$^{58}$\BESIIIorcid{0000-0001-6593-5665},
O.~Bakina$^{37}$\BESIIIorcid{0009-0005-0719-7461},
Y.~Ban$^{47,g}$\BESIIIorcid{0000-0002-1912-0374}
% ...
\\
\vspace{0.2cm}
(BESIII Collaboration)\\
\vspace{0.2cm} {\it
$^{1}$ Institute of High Energy Physics, Beijing 100049, People's Republic of China\\
% ...
$^{4}$ Budker Institute of Nuclear Physics SB RAS (BINP), Novosibirsk 630090, Russia\\
% ...
$^{36}$ Johannes Gutenberg University of Mainz, Johann-Joachim-Becher-Weg 45, D-55099 Mainz, Germany\\
$^{37}$ Joint Institute for Nuclear Research, 141980 Dubna, Moscow region, Russia\\
% ...
$^{47}$ Peking University, Beijing 100871, People's Republic of China\\
% ...
$^{58}$ Southeast University, Nanjing 211100, People's Republic of China\\
$^{59}$ State Key Laboratory of Particle Detection and Electronics, Beijing 100049, Hefei 230026, People's Republic of China\\
% ...
$^{73}$ University of Science and Technology of China, Hefei 230026, People's Republic of China\\
% ...
$^{76}$ University of Turin and INFN, (A)University of Turin, I-10125, Turin, Italy; (B)University of Eastern Piedmont, I-15121, Alessandria, Italy; (C)INFN, I-10125, Turin, Italy\\
$^{77}$ Uppsala University, Box 516, SE-75120 Uppsala, Sweden\\
% ...
$^{82}$ Zhengzhou University, Zhengzhou 450001, People's Republic of China\\
\vspace{0.2cm}
% ...
$^{b}$ Also at the Moscow Institute of Physics and Technology, Moscow 141700, Russia\\
% ...
$^{g}$ Also at Key Laboratory of Nuclear Physics and Ion-beam Application (MOE) and Institute of Modern Physics, Fudan University, Shanghai 200443, People's Republic of China\\
% ...
}      
\end{center}
\vspace{0.4cm}
\end{small}
}
\noaffiliation{}